\documentclass[12pt]{iopart}
\usepackage[utf8]{inputenc}
\usepackage[square,numbers]{natbib}
\usepackage{graphicx}
\usepackage{subfigure}
\usepackage{siunitx}
\usepackage{placeins}
\usepackage{color}
\usepackage{standalone}
\usepackage{dcolumn}
\usepackage{bm}
\usepackage{microtype}
\usepackage{etoolbox}
\usepackage{amssymb}
\usepackage{accents}
\usepackage[normalem]{ulem}
\usepackage[dvipsnames]{xcolor}
\usepackage[colorlinks,urlcolor=NavyBlue,citecolor=NavyBlue,linkcolor=NavyBlue,pdfusetitle]{hyperref}
\usepackage[all]{hypcap}
\usepackage{enumerate}
\usepackage{enumitem}
\usepackage{gensymb}
\usepackage{comment}

\begin{document}

\title[Black Hole Supernovae]{Gravitational waves from core-collapse supernovae with no electromagnetic counterparts}
\author{Jade Powell$^{1,2,*}$, Bernhard M\"uller$^{3}$ }

\address{$^{1}$ Centre for Astrophysics and Supercomputing, Swinburne University of Technology, Hawthorn, VIC 3122, Australia}
\address{$^{2}$ OzGrav: The ARC Centre of Excellence for Gravitational-wave Discovery, VIC 3122, Australia}
\address{$^{3}$ School of Physics and Astronomy, Monash University, Clayton, VIC 3800, Australia}
\address{$^*$Author to whom any correspondence should be addressed.}
\ead{dr.jade.powell@gmail.com}
\vspace{10pt}
\begin{indented}
\item[]\today
\end{indented}


\begin{abstract}
Core-collapse supernovae (CCSNe) are regularly observed electromagnetically, prompting targetted searches for their gravitational-wave emission. However, there are scenarios where these powerful explosions may not have any observable electromagnetic signal, but would still have strong detectable emission in gravitational waves and neutrinos. A regular CCSN explosion may be obscured by matter in the Galaxy. A star may undergo a failed CCSN explosion, where the stalled shockwave is not revived, and would eventually form a black hole. Higher mass progenitor stars may revive the shock, but form a black hole too quickly for the shockwave to reach the surface of the star and produce an electromagnetic signal. Previous work has shown that we can determine if a black hole forms from the CCSN neutrino emission if there are long-duration sinusoidal modulations in the neutrino signal caused by the standing accretion shock instability (SASI). The SASI also produces an observable signature  in the gravitational-wave emission. In this paper, we investigate if we can distinguish between different scenarios for electromagnetically dark CCSNe using the gravitational-wave emission alone. We find, using a reconstruction of the SASI mode, abrupt end times of the gravitational-wave emission, and the rate of change of frequency of the dominant mode, that we are able to accurately distinguish between an obscured CCSN, a failed CCSN, and an explosion with fast black hole formation. 
\end{abstract}

\section{Introduction}
\label{sec:intro}

Core-collapse supernovae (CCSNe) are among the most luminous electromagnetic (EM) transients observed in the Universe. They are also powerful emitters of gravitational waves and neutrinos, and will likely result in the first ever joint gravitational-wave, neutrino and EM detection. The first direct detection of gravitational waves was made 10 years ago, from the merger of two black holes in a binary system \cite{GW150914}. Since the first discovery, the gravitational-wave detectors Advanced LIGO \cite{aLIGO_15}, Advanced Virgo \cite{Virgo_15}, and KAGRA \cite{Kagra_20} have increased in sensitivity, and hundreds of gravitational-wave signals from binary black hole and neutron star mergers have now been discovered \cite{gwtc3}. This number will soon grow to thousands as the gravitational-wave detectors further increase in sensitivity over the coming years. One of the next major goals of gravitational-wave science will then be to detect bursts of gravitational waves from new astronomical sources, such as CCSNe \cite{Powell_24b}. 

CCSNe occur in stars with zero-age main sequence (ZAMS) masses larger than about $8\,\mathrm{M}_{\odot}$. Such massive stars undergo several stages of thermonuclear fusion, up to iron, and then undergo gravitational collapse when the iron core reaches the effective Chandrasekhar mass. The core of star then \textit{bounces}, launching a shock wave that quickly loses energy due to photodisintegration and neutrino losses, and thus stalls at $\sim 100$\,km, as a proto-neutron star (PNS) forms at the center. The shock wave then needs to gain more energy to power an actual CCSN explosion. Regular CCSNe are thought to explode by absorbing some of the energy from the emitted neutrinos \cite{janka_17}. More extreme CCSNe may have explosions powered by rapid rotation and strong magnetic fields \cite{moesta_14,reichert_23,mueller_25}. As EM observations cannot provide us with direct information about the inner regions of the exploding star, a gravitational-wave detection is needed for us to fully understand the mechanisms powering CCSN explosions.  CCSNe are observed regularly through EM telescopes, and one (SN\,1987A) has been observed by neutrino detectors \cite{hirata_87}, but they have not yet been observed through gravitational waves \cite{sn_search_o1o2, szczepanczyk_24, sn2023ixf}. 

Predictions of gravitational-wave signals from CCSNe come from computationally expensive numerical simulations. The modelling of gravitational waves from CCSNe has advanced rapidly in the 10 years since the first detection of gravitational waves. At the time of the first gravitational-wave detection, there were only a very small number of gravitational-wave predictions from 3D simulations of CCSNe \cite{scheidegger_10, mueller_12, ott_13}. The few available waveforms from 3D simulations had issues such as only covering the first few hundred milli-seconds of the gravitational-wave emission, or inadequate neutrino transport. A much larger number of 2D CCSN gravitational-wave predictions existed \cite{dimmelmeier_08, murphy_09,mueller_13,abdikamalov_14, yakunin_15}, but 2D simulations tend to overestimate the gravitational-wave amplitude, and they can only produce one of the two gravitational-wave polarizations. This made it difficult to produce accurate estimates of the maximum CCSN detection distance for the LIGO-Virgo-KAGRA observatory network  \cite{gossan_16}, and to develop accurate data analysis algorithms for determining the astrophysical properties of the source \cite{Powell_16}. 

In the 10 years since the first gravitational-wave detection, there has been a significant number of publications of 3D gravitational-wave signals for non-rotating CCSNe \cite{andresen_17, kuroda_17, yakunin_17, oconnor_18, radice_19, Powell_19, Powell_20, Powell_21, mezzacappa_23, burrows_24}. Although the amplitude of the gravitational waves can vary between different simulation codes, the time-frequency morphology of the signal is now very well understood. Spectrograms of the gravitational-wave signals show a dominant g/f-mode that rises in gravitational-wave frequency with time as the PNS gains mass and shrinks in radius. There may also be a lower frequency mode due to the standing accretion shock instability (SASI) \cite{blondin_03}, a low-mode
(dipolar or quadrupolar) oscillatory instability of the supernova shock that is predicted to occur when fast post-shock advection suppresses neutrino-driven convection \citep{foglizzo_15}. The SASI mode increases in gravitational-wave frequency with time, up to the shock revival time, after which it either disappears or produces much lower frequency gravitational waves as the shock rapidly expands. Recent work has shown that in exploding models, there can also be very high amplitude gravitational-wave emission below 10\,Hz, mainly due to the asymmetric emission of neutrinos \cite{mueller_04,vartanyan_23, Powell_24a}. This aspect of the gravitational-wave emission is a promising source for space-based gravitational-wave detectors \cite{moon_25}. However, longer duration simulations are still needed to fully understand the science case for detecting low frequency gravitational waves from CCSNe. 

In the case of rotation and magnetic fields, only the spike in the gravitational-wave time series that occurs at core bounce was well understood 10 years ago \cite{dimmelmeier_08, abdikamalov_14, scheidegger_10}, but there were no long duration gravitational-wave signals beyond the post-bounce phase. In recent years, the CCSN community has now produced a small number of gravitational waveforms for more extreme CCSNe with rapid rotation and strong magnetic fields, simulated in 3D \cite{bugli_20, obergaulinger_20, Powell_23, Powell_24a}. Rotation and magnetic fields often lead to strong, early explosions without any lower frequency gravitational waves from the SASI. The dominant g/f-mode in the gravitational-wave signal looks similar to the non-rotating case, but with significantly higher gravitational-wave amplitudes, leading to larger detection distances. 

The availability of significantly more advanced gravitational-wave signals for CCSNe has enabled the development of better data analysis tools for their detection and parameter estimation. Several studies have used these numerical waveforms to create models for different CCSN explosion mechanisms \cite{heng_09, rover_09, logue_12, Powell_16, suvorova_19, roma_19, Powell_23b}. Others have developed tools for measuring the rotation rate and the equation of state \cite{chao_22, pastor_24}, and the dominant g/f-mode \cite{bruel_23}. Work has also begun on developing phenomenological waveform models for CCSNe \cite{cerda_duran_25}, as it will never be possible to predict the exact gravitational-wave signal for every CCSN progenitor star, due to their stochastic nature, and the computational expense of 3D simulations. The gravitational-wave signal can also be reconstructed with model-agnostic techniques, such as a wavelet analysis \cite{szczepanczyk_21, raza_22, hu_22, yuan_24}. However, these works do not go into any detail about how their waveform reconstructions relate to the underlying astrophysics of the source.   

The current LIGO-Virgo-KAGRA searches for gravitational waves from CCSNe use search algorithms that make minimal assumptions about the morphology of a gravitational-wave signal \cite{sn_search_o1o2, szczepanczyk_24, sn2023ixf}. The searches target specific CCSNe that have been observed in EM during gravitational-wave detector observing runs \cite{sn_search_o1o2, szczepanczyk_24, sn2023ixf}. The closest CCSN whilst a gravitational-wave detector was operational was SN\,2023ixf at $\sim 6$\,Mpc, but no gravitational-wave signal was detected \cite{sn2023ixf}. Unfortunately, the gravitational-wave detectors were offline for a significant fraction of the time window where it is estimated that the gravitational-wave signal would have occurred. However, if the gravitational-wave signal did occur whilst the gravitational-wave detectors were online, then the constraints on the gravitational-wave amplitude can be improved by a factor of 10 in comparison to the targetted gravitational-wave searches in earlier observing runs. 

The advances in gravitational waveform modelling for CCSNe has enabled groups to start development work on CCSN specific gravitational-wave searches that target the known time-frequency morphology of the signal \cite{astone_18}. 
However, detection rates for current gravitational-wave detectors are only a few every hundred years, as maximum detection distances are Galactic for typical CCSNe, or  out to the Magellanic Clouds for more extreme CCSN models \cite{szczepanczyk_21}. Next-generation gravitational-wave detectors will have a much larger distance reach, and may detect a CCSN every few years.

There are also model-independent gravitational-wave burst searches that search the entire sky for gravitational-wave events \cite{lvk_allsky_O2, lvk_O3_allsky, lvk_O4a_allsky}. They can produce reconstructions of the gravitational-wave signal, and measurements of properties like the central frequency, the gravitational-wave amplitude and the duration. However, testing these all-sky searches on the latest available 3D CCSN waveforms has shown that they have a lot of room for improvement in their sensitivity to CCSN signals. The latest all-sky gravitational-wave search, from the first part of the fourth observing run, was only sensitive to the most extreme CCSNe at distances beyond the galactic center \cite{lvk_O4a_allsky}. This type of all-sky search for CCSN gravitational-wave emission is important, because not every CCSN with detectable gravitational-wave emission will be EM bright. Many CCSNe, about 10-30\%, are thought to be obscured by gas and dust in the Galaxy \cite{Fox_21}. The last supernova observed in our own Galaxy was in 1604 \cite{vink_17}. However, there have been observations of younger Galactic CCSN remnants \cite{reynolds_08}. As gravitational waves and neutrinos couple weakly with matter, these dust-obscured CCSNe would look like regular CCSNe in gravitational-wave and neutrinos detectors.  

Recent 3D CCSN simulations have shown that some CCSNe do not undergo shock revival, resulting in no EM counterpart, and likely black hole formation at a later time \cite{oconnor_18, burrows_24}. These CCSNe are still strong emitters of gravitational waves. The disappearance of the progenitor star could potentially be observed by EM observatories, after the emission of the gravitational waves and neutrinos \cite{adams_17}. Simulations have shown that these non-exploding models have long duration SASI modes in their gravitational-wave emission \cite{burrows_24}. The SASI also produces a long and strong signature in the neutrino emission \citep{tamborra_14, mueller_19}. Previous work has shown that we can determine if the CCSN forms a black hole by detecting this SASI signature in the neutrino emission \cite{walk_20}.
In fact, the gravitational-wave and neutrino signal from the SASI would provide the first direct evidence that the instability occurs in supernova cores.
Even though there is currently no observational confirmation of the SASI, there is significant observational evidence on low-mode dipole asymmetries in supernova explosions, e.g., from polarisation signatures and supernova remnant observations \cite{foglizzo_15, grefenstette_17, katsuda_18}. The instability has also been shown to occur for strongly retracted shocks from shallow-water experiments, which provide a close analogue of the accretion flow onto a PNS \cite{foglizzo_12}. 

Some CCSNe with massive progenitor stars form black holes only a few hundred milli-seconds after the core bounce time. Recent work has shown that massive stars, greater than $\sim 35\,\mathrm{M}_{\odot}$, often undergo shock revival before black hole formation \cite{chan_18,chan_20b,Powell_21, pan_21, burrow_24b}. 
These models have recently been referred to as \textit{BHSNe}. If the revived shock is weak, or the time between the shock revival and black hole formation is only a few hundred milli-seconds, there will likely be no EM bright explosion. However, there are some 3D simulations with very powerful explosions a second or two before black hole formation, that would still result in very EM bright CCSNe \cite{burrow_24b}. 
Although there is some evidence that red supergiants above
$18\texttt{-}20\,\mathrm{M}_\odot$ do no explode\footnote{The statistical significance of the constraints on red supergiant explosions continues to be analysed and discussed in the literature in the light of uncertainties in the determination of progenitor masses \citep{beasor_25}.}
\citep{smartt_15}, a small number of black-hole forming explosions from more massive stars remains compatible with the observational evidence, and would explain
individual supernovae with very high ejecta mass \citep{terreran_17} and low-mass black holes as seen
in the compact object merger event GW190814 \citep{GW190814_2020, antoniadis_22}.
In the case of quiet black-hole formation, it is possible that there may still be some EM signature, as the star puffs away its outer layers, similar to a luminous red nova \cite{lovegrove_13}. This would not be the case for stars that have been stripped of their outer layers, due to winds or binary interactions, before the CCSN occurs. 

In this paper, we investigate for the first time if the gravitational-wave emission alone can be used to distinguish between obscured CCSNe, non-exploding failed CCSNe, and the BHSNe that undergo shock revival before quickly forming a black hole. In our previous work \cite{Powell_22}, we added an asymmetric chirplet model to the Bayesian parameter estimation code Bilby \cite{bilby_19} to reconstruct the CCSN g/f-mode. As the SASI mode is essential for determining if the shock is revived, in this work, we improve our previous CCSN gravitational-wave signal model so that it also includes a lower frequency SASI mode. We test our improved signal model on 11 different 3D CCSN waveforms from various different simulation codes. We show how well our model can reconstruct the CCSN gravitational-wave signals, and we show how the duration of the g/f-mode, the duration of the SASI mode, and the rate of change of frequency of the g/f-mode can be used to distinguish between the different EM dark CCSN scenarios. 

The outline of the paper is as follows: In Section \ref{sec:waveforms}, we describe the different CCSN waveforms that we use to represent gravitational-wave detections from different EM dark scenarios. In Section \ref{sec:model}, we describe the improvements to our CCSN gravitational-wave signal model. In Section \ref{sec:data}, we describe the simulated data used for this study. In Section \ref{sec:reconstruct}, we show the reconstructed waveforms. In Section \ref{sec:emdark_model}, we show that we can accurately distinguish between the different types of EM dark CCSNe. We discuss the maximum detectable distances for the various CCSN gravitational-wave models in Section \ref{sec:dist}. A discussion and conclusions are given in Section \ref{sec:conclusions}.

\section{Supernova Models}
\label{sec:waveforms}

We select 11 different waveforms to represent gravitational-wave detections from three different types of CCSNe that may not have an EM counterpart. All of the waveforms used here are from 3D simulations using a variety of different CCSN simulation codes. 

\subsection{Obscured CCSNe}


\begin{figure}
\includegraphics[width=\textwidth]{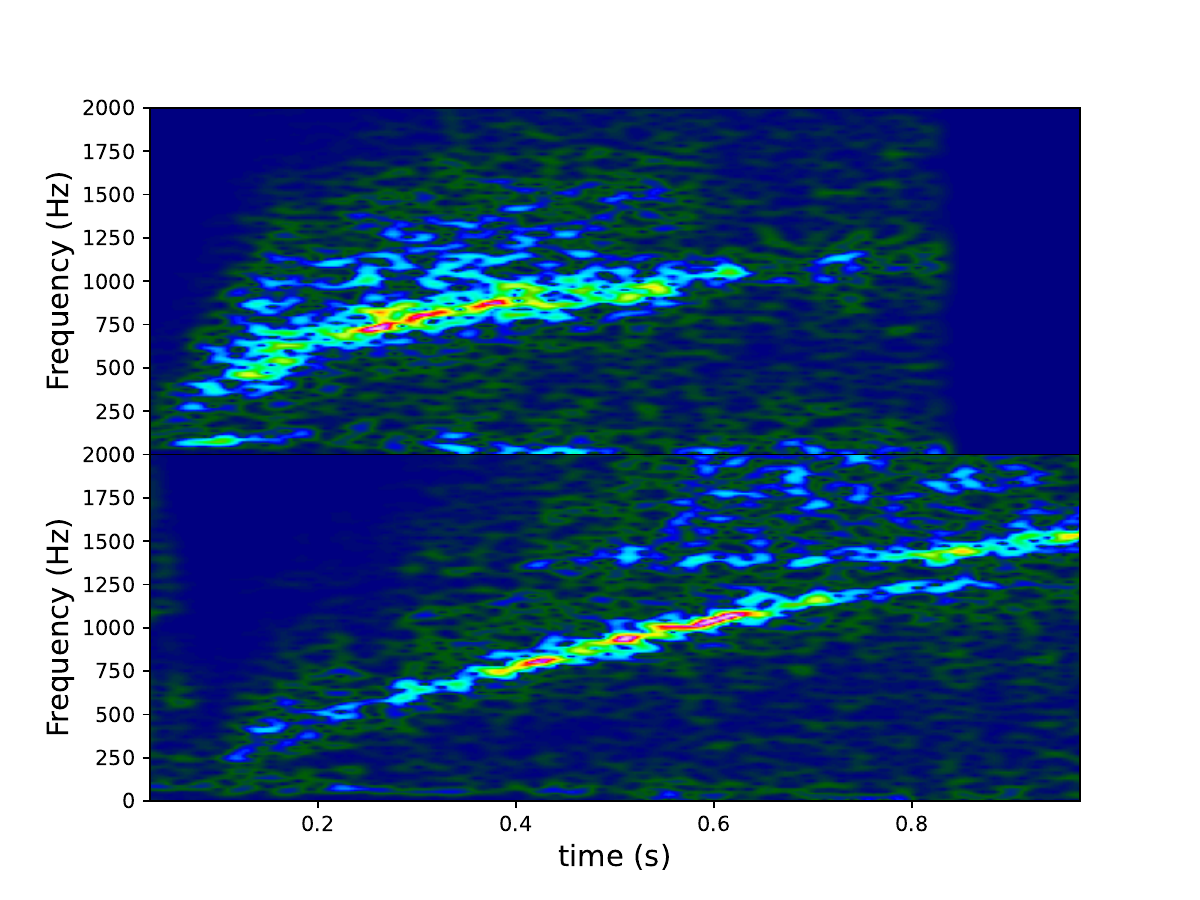}
\caption{Examples of gravitational-wave emission from typical neutrino-driven explosions. Regular CCSNe may not be observed electromagnetically if they are obscured by matter in the Galaxy. The top panel shows model s18 from \cite{Powell_19}. The bottom panel shows model s24 from \cite{burrows_24}. The main feature in the gravitational-wave emission is the dominant g/f-mode. A SASI mode may also be present before the time of shock revival.}
\label{fig:neutrino}
\end{figure}

It is possible that a regular CCSN may occur in our own Galaxy, but the EM signal may be entirely obscured by other objects like the Galactic Centre. However, the gravitational-wave signal will easily pass through the matter blocking the EM light, and we would expect the gravitational-wave emission to look the same as it would for a CCSN with an EM counterpart. 

Regular neutrino-driven explosions have now been simulated extensively in 3D, providing us with many example gravitational-wave signals from across the progenitor mass range. Some examples of the gravitational-wave signals used here are shown in Figure \ref{fig:neutrino}. The main feature of the waveforms is the dominant high-frequency g/f-mode. The g/f-mode can be up to several seconds long, but should be a minimum of $\sim0.5$\,s before the gravitational-wave amplitude becomes too low for detection. They may also have a lower frequency SASI mode before the revival of the shock.  

Here we use four different progenitor models. The first is model s18, a solar metallicity $18\,\mathrm{M}_{\odot}$ ZAMS mass progenitor from \cite{Powell_19}. The model undergoes shock revival at $\sim 250$\,ms after core bounce. The second is model he3.5 from \cite{Powell_19}. This model is a $3.5\,\mathrm{M}_{\odot}$ helium core at the time of the core-collapse, as it has been stripped of its outer layers due to binary interactions. It undergoes shock revival at $\sim 450$\,ms after core bounce, resulting in some visible low frequency gravitational-wave emission before this time. The signal ends at 0.7\,s after core-bounce. 
The third model in this group is the $20\,\mathrm{M}_{\odot}$ y20 model from \cite{Powell_20}. The shock is revived about $200\,$ms after core bounce, and the gravitational-wave frequencies reach about 1000\,Hz before the simulation end at over 1\,s. The fourth model is the $24\,\mathrm{M}_{\odot}$ ZAMS mass model s24 from \cite{burrows_24}, which also undergoes shock revival about $200$\,ms after core bounce. The s24 model has a visible dominant mode for several seconds, however we only use the first second of the gravitational-wave emission. The dominant mode reaches a frequency of $\sim 1800$\,Hz at 1\,s. 

\subsection{No Explosion}


\begin{figure}
\includegraphics[width=\textwidth]{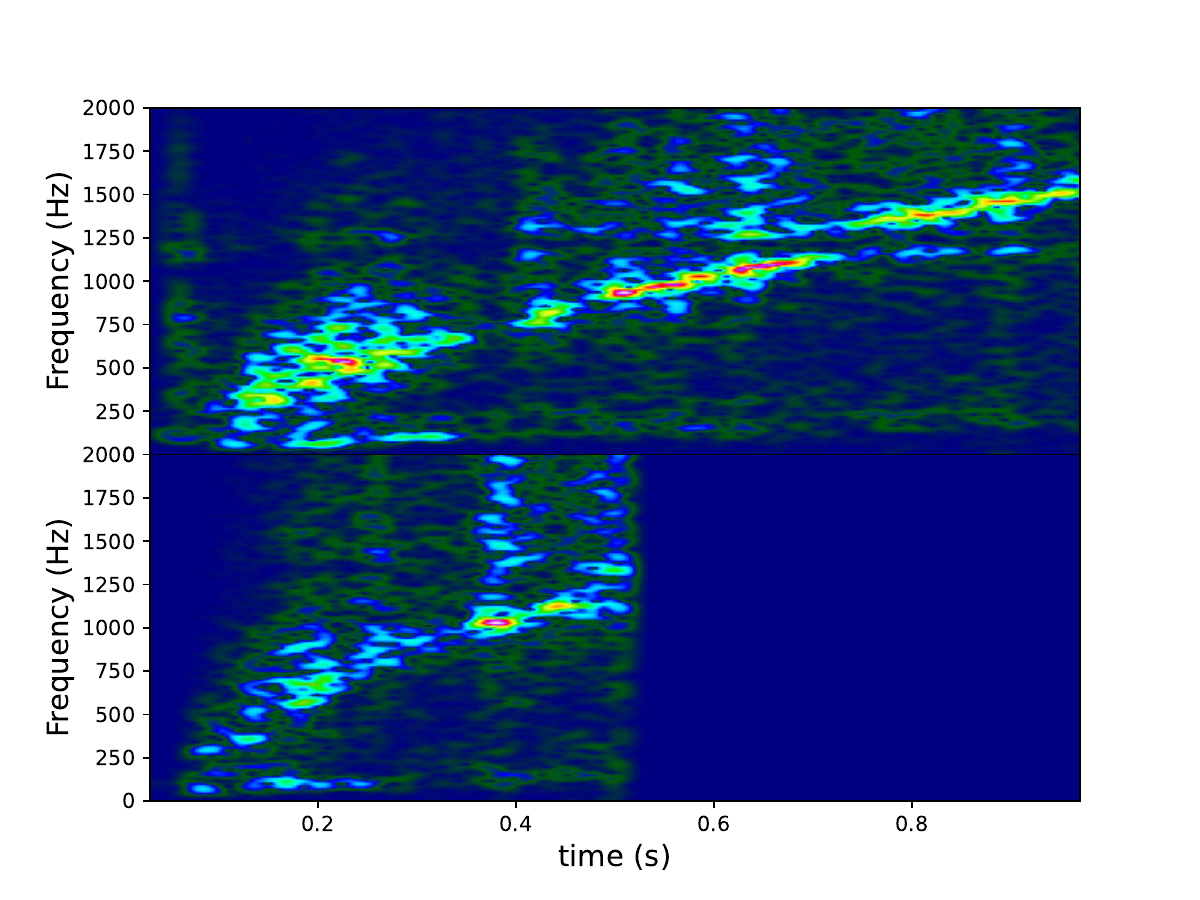}
\caption{Examples of gravitational-wave emission from CCSNe that do not undergo shock revival. Here we show model s14 (top) from \cite{burrows_24}, and model mesa20\_pert from \cite{oconnor_18}. The models have a dominant high frequency f/g-mode that is typical for all CCSN gravitational-wave signals. As the models do not undergo shock revival, they also have a lower frequency SASI mode that continues to increase in frequency with time for the entire duration of the signal.  }
\label{fig:no_exp}
\end{figure}

The next group of CCSN waveforms considered are models that do not undergo shock revival during the simulation time, and would most likely silently form a black hole at much later times. Examples are shown in Figure \ref{fig:no_exp}. Models that do not undergo shock revival tend to have lower gravitational-wave amplitudes. They have a strong lower frequency SASI mode in their gravitational-wave emission, due to much longer periods of SASI activity than models that fully explode. The frequency of the SASI is well approximated by 
\begin{equation}
f_\mathrm{SASI} \propto R_\mathrm{sh}^{-3/2} ,
\label{eqn:sasi}
\end{equation} 
where $R_\mathrm{sh}$ is the shock radius \cite{mueller_14}. When the shock is stalled, the SASI mode slowly increases in frequency with time. After shock revival, the SASI mode frequency rapidly drops to below the LIGO-Virgo-KAGRA frequency band, or the SASI activity is killed by the shock revival. As the non-exploding models do not undergo shock revival, their SASI mode continues to increase in frequency with time until the end of the simulation.  

We use model s12.25, which has a $12.25\,\mathrm{M}_{\odot}$ ZAMS mass progenitor star, and model s14, which has a $14\,\mathrm{M}_{\odot}$ ZAMS mass progenitor. Both of the models are from \cite{burrows_24}. The two models are $\sim2$\,s long, and reach frequencies of over 2000\,Hz. The amplitude of the gravitational waves stay high for the entire duration of the simulation. They have a clear low frequency SASI mode, that rises in frequency with time. For this work, we only consider the gravitational-wave emission as calculated in the `y-direction'. 

The third model we use is the $20\,\mathrm{M}_{\odot}$ ZAMS mass progenitor model mesa20\_pert from \cite{oconnor_18}, which we hereafter refer to as model m20. The dominant high-frequency g/f-mode reaches frequencies of about 1250\,Hz, and the simulation is stopped after about half a second, whilst the gravitational-wave amplitude is still high. This model also shows a low frequency SASI mode that increases in frequency with time up to the end of the simulation. 

The fourth non-exploding model we use is model s18np from \cite{Powell_20}. This model has the same $18\,\mathrm{M}_{\odot}$ progenitor as the s18 model, but model s18np does not include any perturbations in the progenitor which are essential for shock revival. This results in an unusually short g-mode, that only reaches frequencies of $\sim 750$\,Hz. The SASI mode is very strong, and is visible right until the simulation end time. 

\subsection{BHSNe}


\begin{figure}
\includegraphics[width=\textwidth]{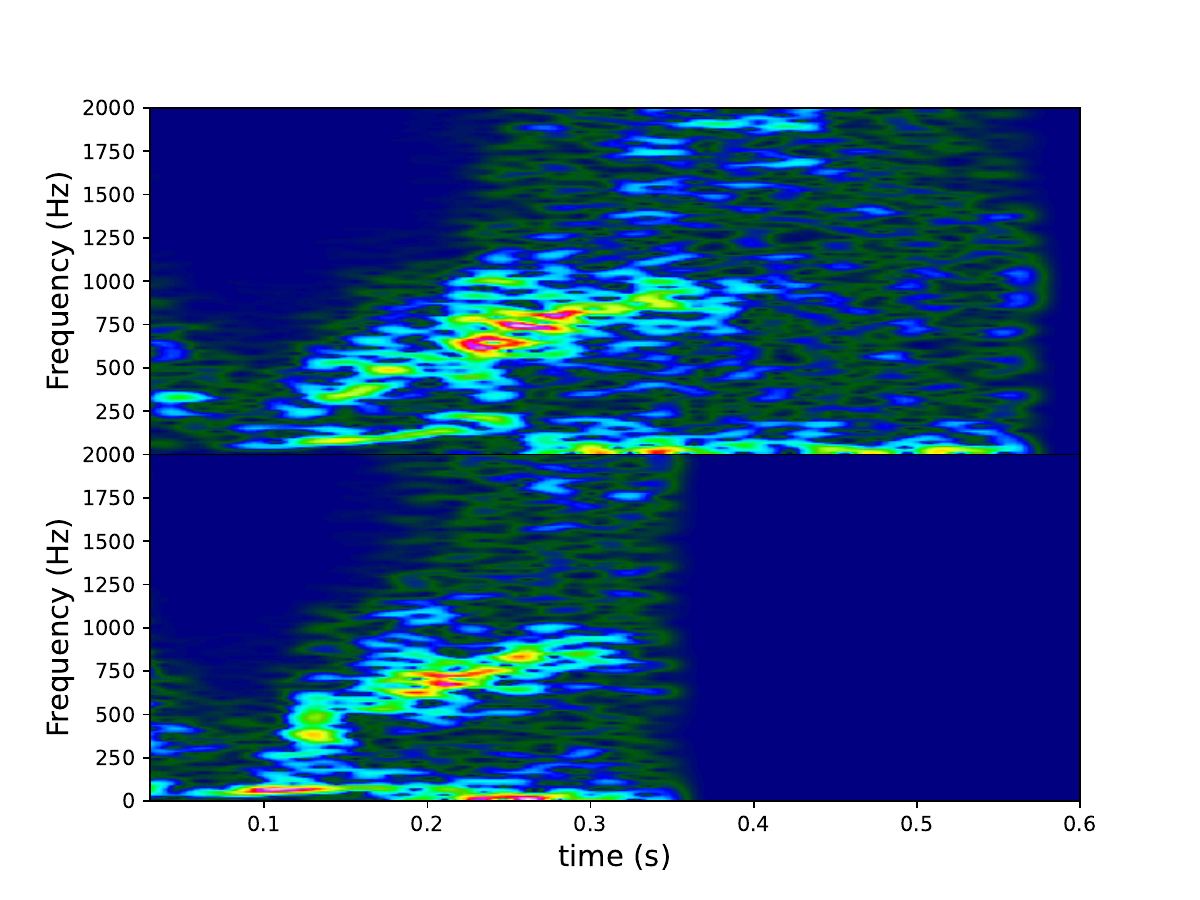}
\caption{Examples of gravitational-wave emission from BHSNe. The top panel shows model z85\_SFHx, and the bottom panel shows model z85\_SFHo, both from \cite{Powell_21}. The models undergo shock revival before quickly forming a black hole. The lower frequency SASI mode rises in frequency with time up until the shock revival. The shock revival, and the generally higher mass progenitor stars, results in high amplitude gravitational-wave emission. }
\label{fig:bhsne}
\end{figure}

This group of waveforms are from stars that collapse and undergo shock revival followed by the formation of a black hole a few hundred milli-seconds later. These \textit{BHSNe} are thought to form a black hole too fast after shock revival to result in a full EM CCSN. As the gravitational-wave emission starts shortly after core bounce, there should be at least a few hundred milli-seconds of gravitational-wave emission before the formation of the black hole. The waveforms used here end abruptly once the black hole forms. In reality, there may be additional features in the gravitational-wave signal at the black hole formation time. There is likely an additional large spike in the gravitational-wave amplitude at black hole formation, followed by some ringdown \cite{uchida_19}. However, current CCSN simulation codes can only calculate the gravitational-wave emission up to right before the black hole formation. The high mass required for rapid black hole formation, and the revival of the shock, generally result in high gravitational-wave amplitudes. 

We use models z85\_LS220, z85\_SFHx, and z85\_SFHo from \cite{Powell_21}. They are $85\,\mathrm{M}_{\odot}$ progenitor stars, at the lower end of the pulsational pair instability supernova region, with three different equations of state. All three models quickly undergo shock revival before forming a black hole a few hundred milli-seconds later. Even though these models undergo energetic explosions, the ratio of the binding energy compared to the explosion energy shows that it is unlikely that the shock ever reaches the surface of the star. 

Two of the gravitational-wave signals are shown in Figure \ref{fig:bhsne}. The gravitational-wave emission ends abruptly once the black hole has formed, and the simulation can no longer continue. The dominant g/f-mode for these models rapidly reaches about 1000\,Hz, and they have a clear lower frequency SASI mode that increases in frequency with time before the shock is revived.

\section{Parameter Estimation Model }
\label{sec:model}

\begin{figure}
\includegraphics[width=\textwidth]{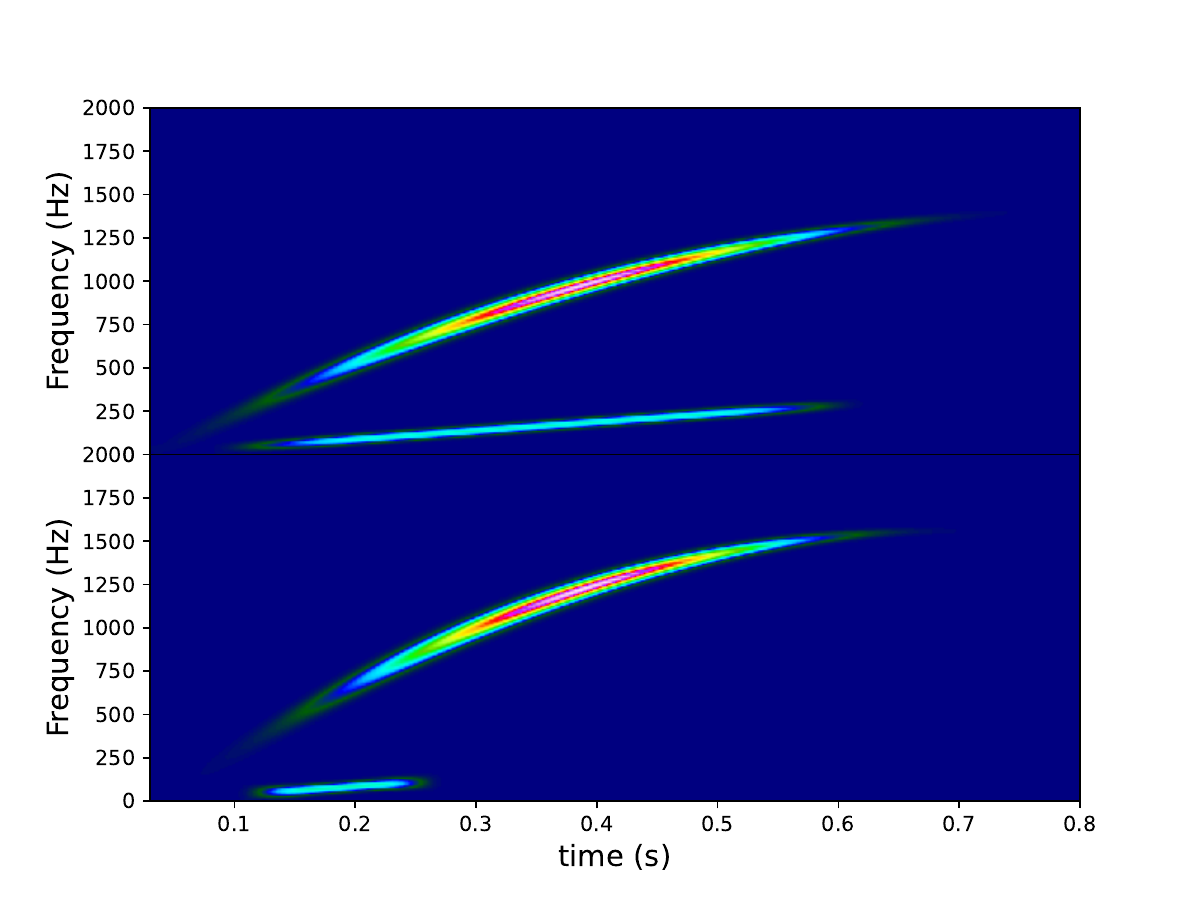}
\caption{Examples of the CCSN signal model for two different sets of parameters. The top panel has a central frequency of 950\,Hz and a SASI duration of 0.55\,s. The bottom panel has a central frequency of 1200\,Hz and a SASI duration of 0.15\,s. }
\label{fig:sig_model}
\end{figure}

To determine if a CCSN detection with no EM counterpart is a BHSNe, non-exploding, or an obscured regular CCSN we use the Bayesian parameter estimation code Bilby \cite{bilby_19}. In our previous CCSN parameter estimation work with Bilby, we developed a new signal model for the analysis of CCSNe \cite{Powell_22}. This signal model represented the dominant g/f-mode emission in CCSN gravitational-wave signals. However, one of the main differences between the three types of CCSN signal models considered here is the lower frequency SASI mode. Therefore, for this work, we extend our previously developed signal model to also include an extra feature to represent the lower frequency SASI mode. 

As in our previous work, we use an asymmetric chirplet model defined in the time domain as,
\begin{equation}
h_{+} = A  \exp\left(-\frac{\delta t^2}{\tau^2}\right) \cos(2\pi f \delta t +  \pi \dot{f} \delta t^2 + c \delta t^3) ,
\end{equation}
\begin{equation}
h_{\times} =  A \exp\left(-\frac{\delta t^2}{\tau^2}\right)  \sin(2\pi f \delta t +  \pi \dot{f} \delta t^2 + c \delta t^3)  , 
\end{equation}
where A is the amplitude, $f$ is the central frequency, $\dot{f}$ is the rate of change of frequency, $\delta t$ is the array of time steps where the central time is the time of maximum gravitational-wave amplitude, and $c$ is a parameter that controls how curved the chirplet is. The parameter $\tau$ is defined at $\tau = Q / 2 \pi f$, where Q is the number of cycles. 
For the new low frequency SASI mode, we add to the signal 
\begin{equation}
h_{+} = A_\mathrm{SASI}   \cos(2\pi f_\mathrm{SASI} \delta t +  \pi \dot{f}_\mathrm{SASI} \delta t^2 ) ,
\end{equation}
\begin{equation}
h_{\times} =  A_\mathrm{SASI}  \sin(2\pi f_\mathrm{SASI} \delta t +  \pi \dot{f}_\mathrm{SASI} \delta t^2)   ,
\end{equation}
where in this work we set $A_\mathrm{SASI}$ to be 20\% of the overall signal amplitude, although this could be made a free parameter in future work. We pick this number based on the amplitude of the SASI mode observed in modern 3D simulations. The parameters $f_\mathrm{SASI}$ and $\dot{f}_\mathrm{SASI}$ are searched for by the parameter estimation code. We set the maximum $f_\mathrm{SASI}$ to 400\,Hz, as this is the maximum value we have observed through 3D CCSN simulations. The start time for the SASI mode we make always equal to the start of the high frequency mode. The duration of the SASI mode is a free parameter searched for by the parameter estimation code. A smooth step function is applied to produce the correct SASI mode length for the given SASI start and end time. For this analysis we set the minimum SASI mode duration to zero, and the maximum to 0.6\,s. We use a flat prior for the SASI duration, $f_\mathrm{SASI}$, $\dot{f}_\mathrm{SASI}$, $Q$, $f$, $\dot{f}$ and $c$. We use a uniform in volume prior for $A$, and uniform on the sky for the sky position. Some examples of our signal model are shown in Figure \ref{fig:sig_model} for a SASI mode duration of 0.15\,s, which represents a model that undergoes shock revival, and 0.55\,s which represents a typical non-exploding model. 

Bilby then uses nested sampling to produce reconstructions of gravitational-wave signals using our new CCSN signal model. We use the non-central chi-squared likelihood described in our previous work, as our CCSN model does not capture the stochastic fluctuations in CCSN signals. The phase isn't required for our study, as only the time-frequency morphology, and the duration of the g/f-mode and SASI mode, are required to understand the CCSN explosion dynamics and the properties of the remnant star. 

\section{Gravitational-Wave Data}
\label{sec:data}

\begin{figure}
\centering
\includegraphics[width=0.9\textwidth]{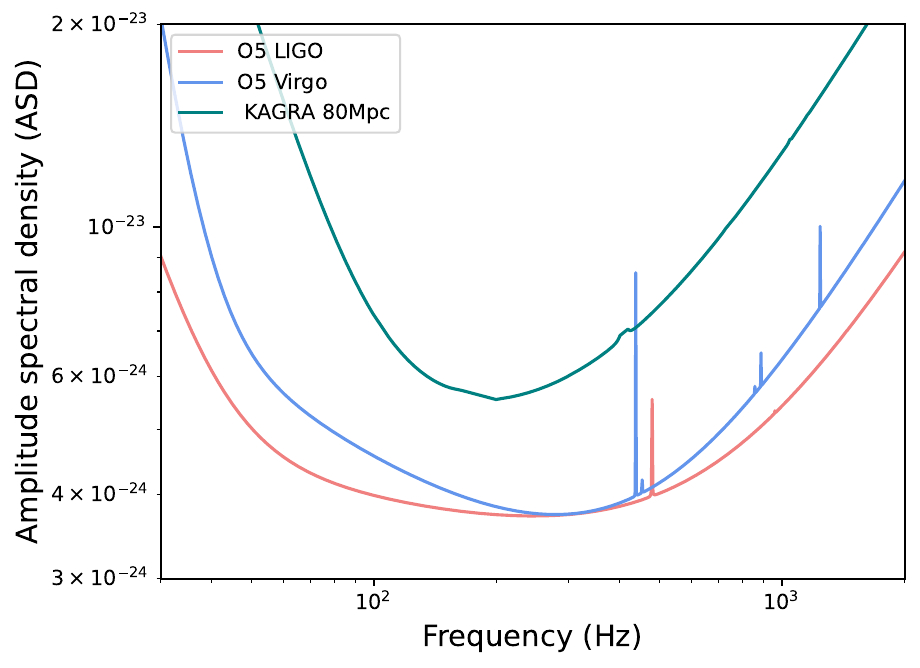}
\caption{The gravitational-wave detector noise curves used in this study. They are the expected detector sensitivities for Advanced LIGO, Advanced Virgo and KAGRA's fifth observing run (O5). The noise curves were downloaded from \cite{dcc, dcc2}. }
\label{fig:noise_curve}
\end{figure}

To understand what we could learn about the source if we detect a CCSN with no EM counterpart during the LIGO-Virgo-KAGRA detector's fifth observing run (O5), we use the predicted O5 detector sensitivity curves downloaded from \cite{dcc} for Virgo and KAGRA, and from \cite{dcc2} for LIGO. The noise curves are shown in Figure \ref{fig:noise_curve}. The detector sensitivity chosen for KAGRA is fairly optimistic with a binary neutron star inspiral detection range of 80\,Mpc. For Virgo, we use the low noise, high range limit predicted noise curve, and for the two LIGO detectors we use the design sensitivity noise curve. 

The detector noise curves are used to produce simulated Gaussian noise for each detector. The supernova waveforms, described in Section \ref{sec:waveforms}, are added to the detector noise at signal to noise ratios (SNRs) between 20 and 50. Below this minimum SNR value, it is unlikely that the current unmodelled gravitational-wave burst searches would be able to see a CCSN signal in O5 \cite{szczepanczyk_24, lvk_O3_allsky}. If new searches that are more sensitive to CCSNe are developed, then we may be able to see CCSN down to much lower SNR values. Some of the current binary black hole gravitational-wave detections have SNR values as low as 7.5 \cite{gwtc3}. 

To prepare the CCSN gravitational-wave signals, we resample the data to 4096\,Hz, and cut off frequencies below 30\,Hz. We only use the first 1\,s of the gravitational-wave signal. For most CCSN models, this covers the highest amplitude phase of the CCSN gravitational-wave emission. We assume that the CCSN signal has already been detected by an all-sky gravitational-wave search, for example \cite{lvk_O3_allsky}. Therefore we assume the time of the gravitational-wave emission is known, even without an EM counterpart, but we do not assume that we know the sky position, as the gravitational-wave sky map could potentially still be large. We then run Bilby to produce reconstructions of our gravitational-wave signals.

\section{Waveform Reconstructions}
\label{sec:reconstruct}

In this section, we show some examples of waveforms reconstructed with our improved CCSN parameter estimation model, and discuss how they can be used to interpret the astrophysical properties of the source. 

\subsection{Obscured CCSNe}

\begin{figure}
\includegraphics[width=0.5\textwidth]{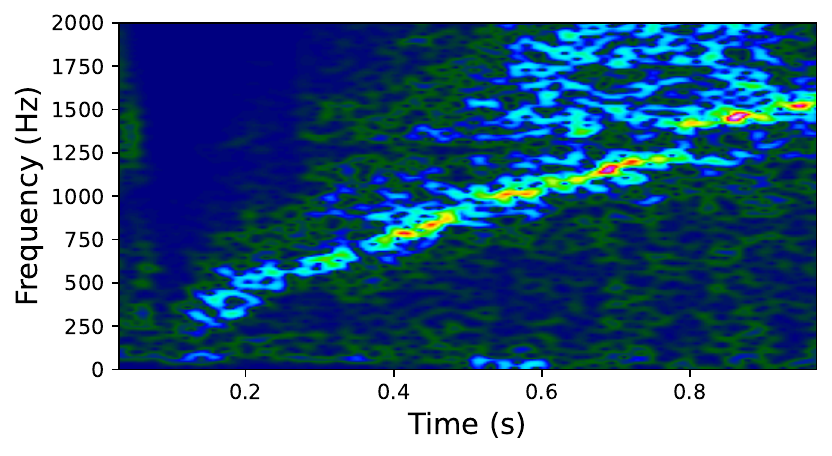}
\includegraphics[width=0.5\textwidth]{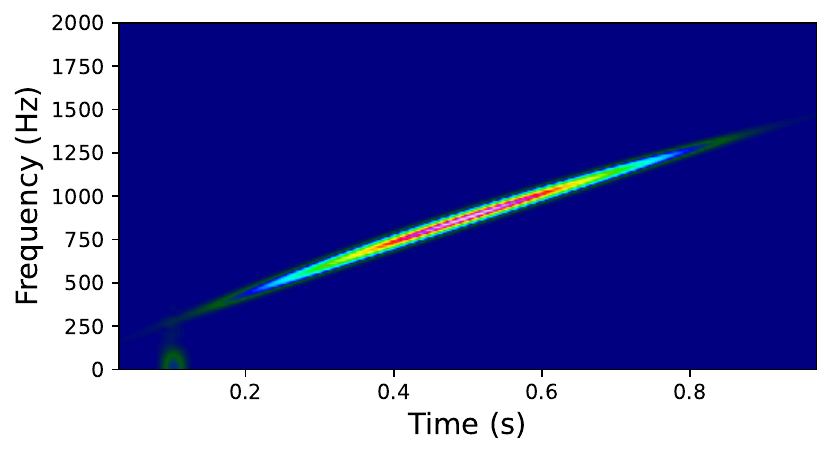}
\includegraphics[width=0.5\textwidth]{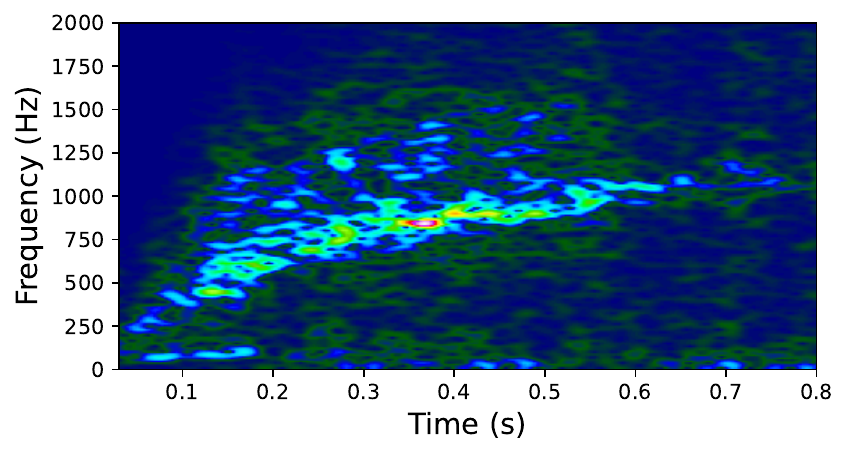}
\includegraphics[width=0.5\textwidth]{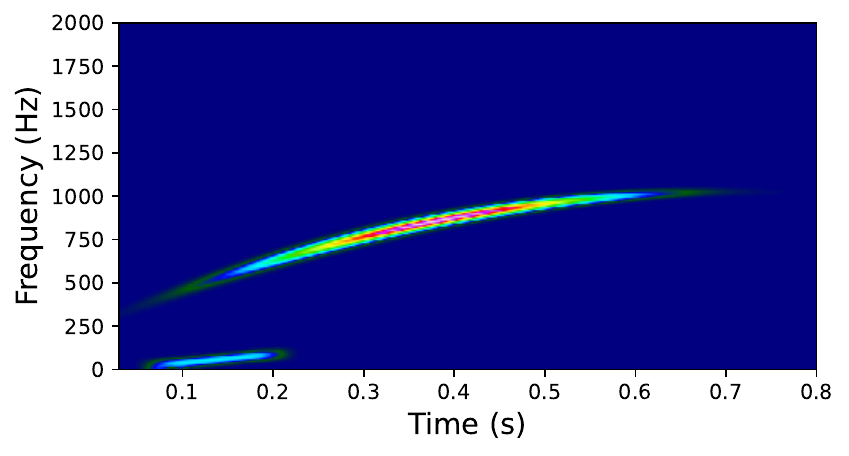}
\includegraphics[width=0.5\textwidth]{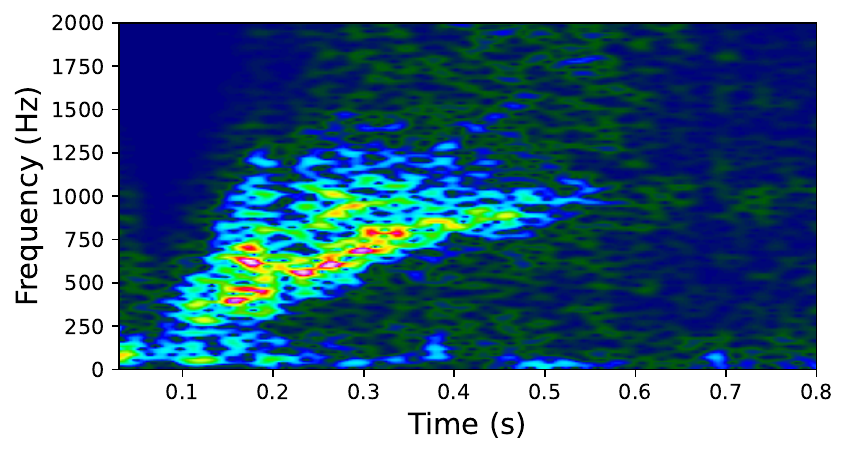}
\includegraphics[width=0.5\textwidth]{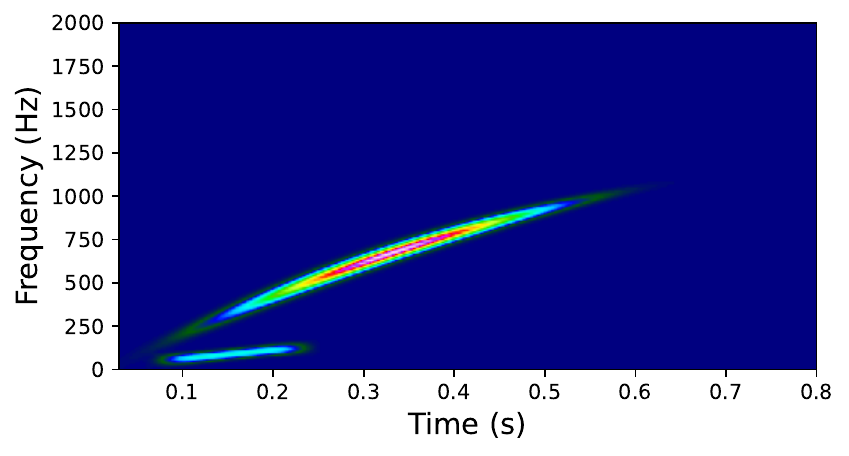}
\includegraphics[width=0.5\textwidth]{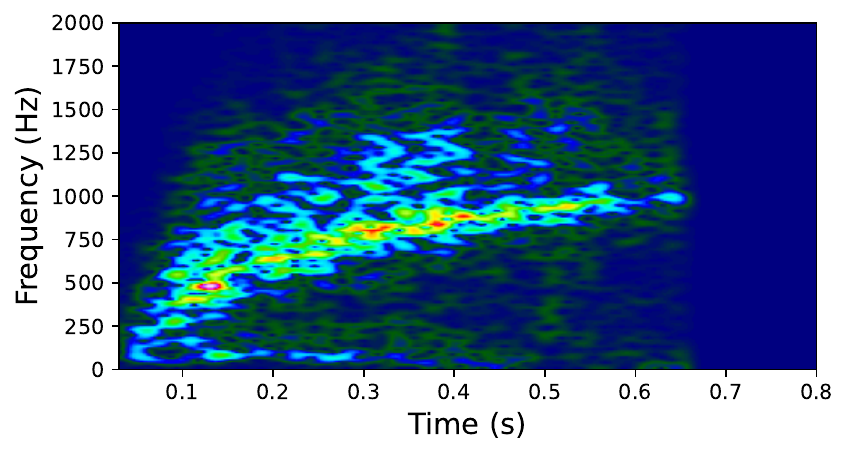}
\includegraphics[width=0.5\textwidth]{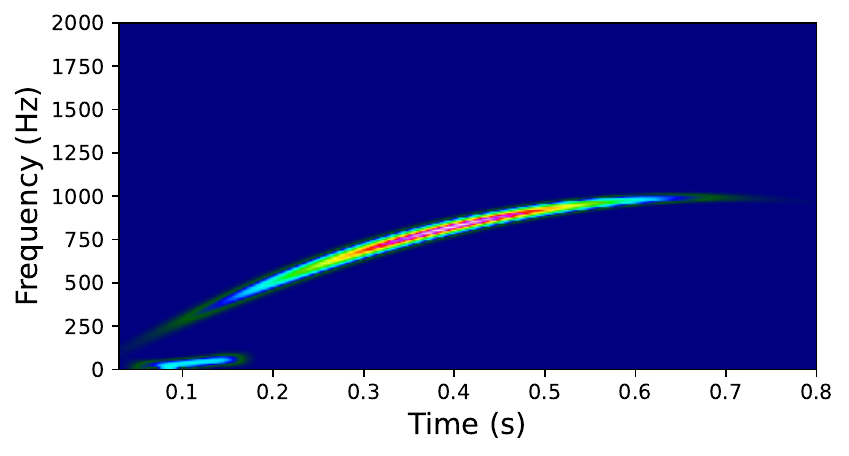}
\caption{Examples of waveform reconstructions for regular CCSNe that may be obscured by matter in the Galaxy. From top to bottom are models s24, s18, y20 and he3.5. The left column is the injected gravitational-wave signal. The right column shows the gravitational-wave signal reconstructed with our model at SNR 40. The models find some low frequency emission from the SASI mode only at very early times in the signals, before the shock revival time. }
\label{fig:expl_results}
\end{figure}

First we consider the case of regular CCSNe that are obscured by the Galaxy. Regular neutrino-driven explosions are the most well understood type of CCSN gravitational-wave signal, as they have now been simulated extensively in 3D. The gravitational-wave signals reconstructed by our model, using the maximum values of the posteriors produced for each model parameter, are shown in Figure \ref{fig:expl_results} at a network SNR of 40. Previous work has shown that other methods for reconstructing CCSN gravitational-wave signals can capture $\sim 70\%$ of the full CCSN signal at an SNR of 40 \cite{szczepanczyk_21, raza_22}. 
However, it is not clear from those previous studies exactly how well that reconstructed 70\% captures the full features of the dominant g/f-mode or the SASI mode.

In our previous work \cite{Powell_22}, we only tested our CCSN signal model on CCSN waveforms from our own 3D simulations with the CoCoNuT code. Model s24 is the first example that shows that we can also reconstruct the g/f-mode from waveforms produced by other groups. We only use the first 1\,s of the gravitational waveform, but the s24 gravitational-wave signal still has detectable gravitational-wave amplitudes beyond this time. This resulted in our measured $Q$ parameter, which determines duration, to rail against the edge of the prior distribution. The overall time-frequency evolution of the g/f-mode is well captured by the signal model, even though the model has a gap due to an avoided crossing. In model s24, there is no reconstructed lower frequency SASI mode, which is an accurate representation of the injected waveform. 

Reconstructing the time-frequency evolution of the g/f-mode is important as the gravitational-wave frequency is determined by the mass and radius of the PNS \cite{Powell_22}. We do not convert our g/f-mode reconstructions here into measurements of the PNS mass and radius, as we demonstrated how accurately this can be done in our previous work \cite{Powell_22}. The reconstruction of the g/f-mode in models s18, y20 and he3.5 matches what we observed in our previous work with the old version of our CCSN signal model \cite{Powell_22}. However, in this work, our new signal model is also able to reconstruct the lower frequency SASI mode. This mode is very short in all of the obscured gravitational-wave signals, as the SASI disappears in these models once they undergo shock revival. At SNR 40, the SASI mode durations are $0.17^{+0.09}_{-0.15}$\,s for model s18, $0.18^{+0.12}_{-0.14}$\,s for model y20, and $0.13^{+0.09}_{-0.04}$\,s for model he3.5. At the lower SNR values we consider here, the reconstructed g/f-modes and SASI modes are similar, but the error bars on the measured parameters become much larger. 

\subsection{Non-Exploding Models}

\begin{figure}
\includegraphics[width=0.5\textwidth]{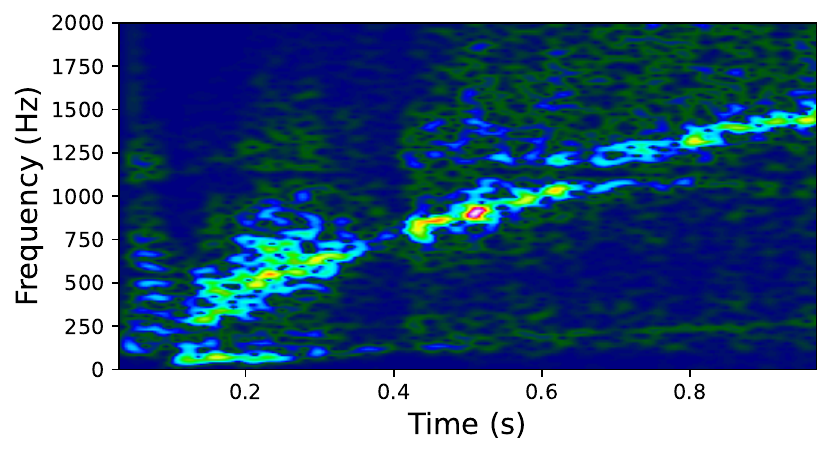}
\includegraphics[width=0.5\textwidth]{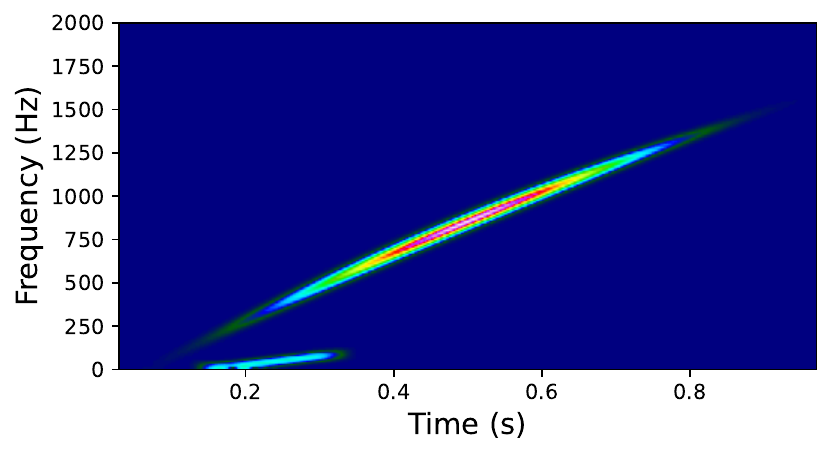}
\includegraphics[width=0.5\textwidth]{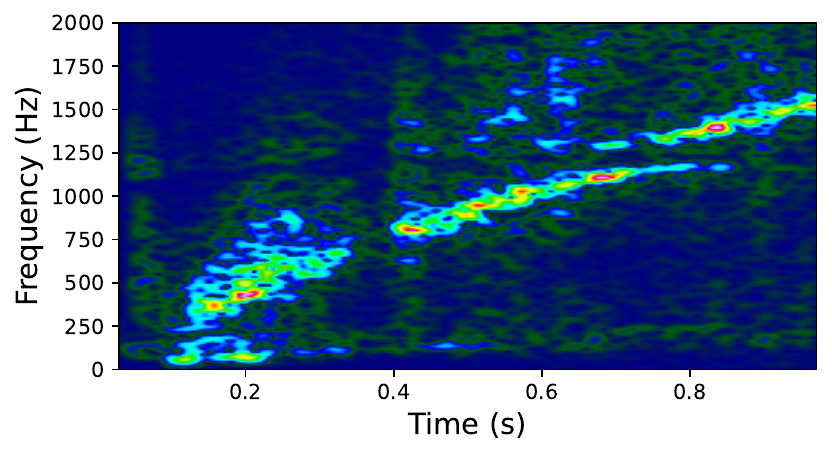}
\includegraphics[width=0.5\textwidth]{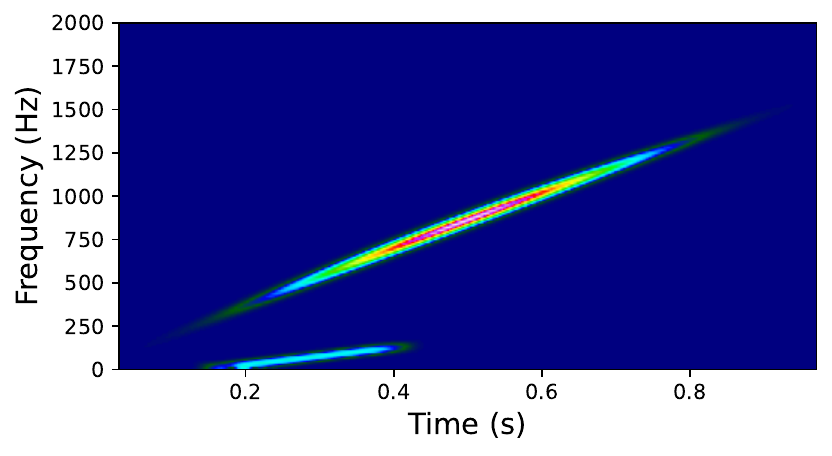}
\includegraphics[width=0.5\textwidth]{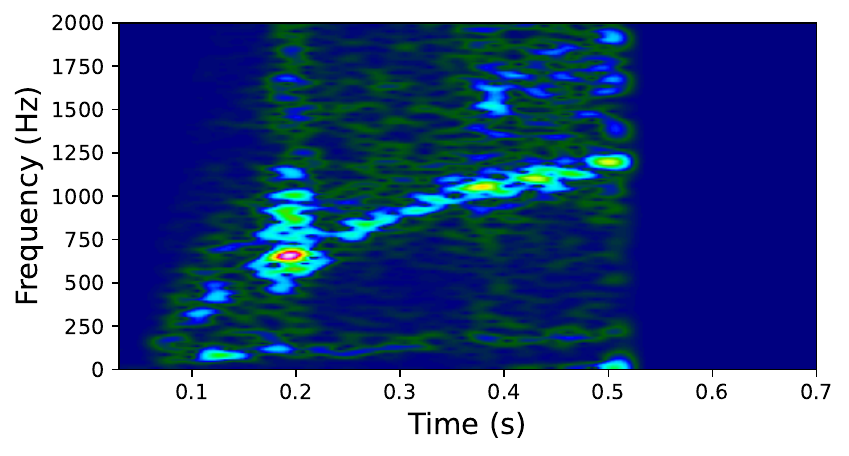}
\includegraphics[width=0.5\textwidth]{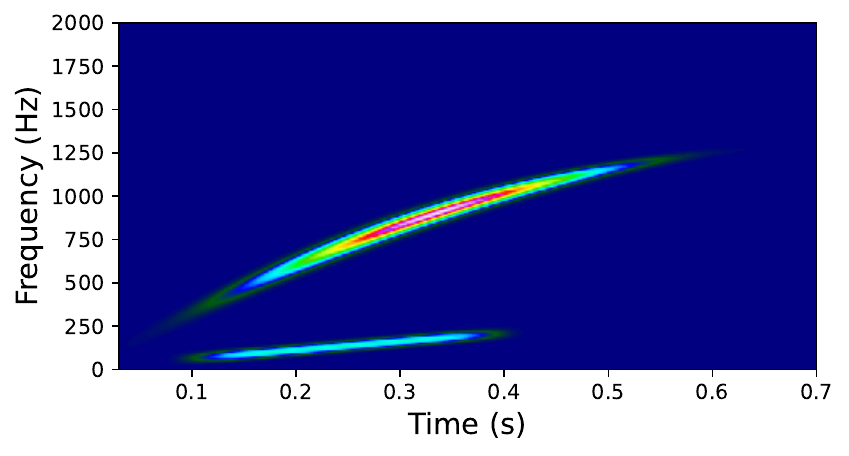}
\includegraphics[width=0.5\textwidth]{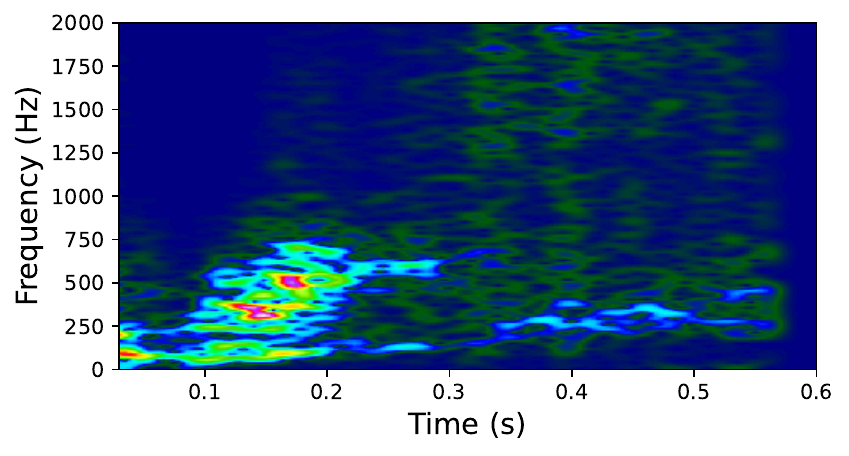}
\includegraphics[width=0.5\textwidth]{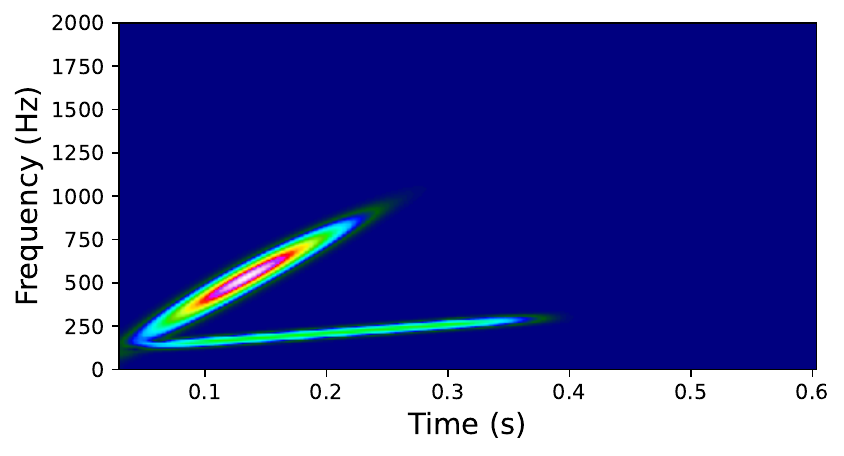}
\caption{Examples of reconstructed gravitational-wave signals from the non-exploding models. From top to bottom are models s12.25, s14, m20 and s18np. The left column shows the injected waveforms. The right column shows the waveforms reconstructed at SNR 40. Gravitational waves from non-exploding CCSNe have a longer SASI mode that increases in frequency with time. }
\label{fig:no_expl_results}
\end{figure}

In Figure \ref{fig:no_expl_results}, we show examples of reconstructed waveforms for the models that do not undergo shock revival and therefore have no EM explosion. They are produced from the waveforms injected at SNR 40, using the peak of the posterior for each of the model parameters. For models s12 and s14, our waveform model is able to reconstruct the higher frequency mode even though there are some gaps in the emission, likely due to avoided crossings. In the injected waveforms, both of these models clearly have a SASI mode that extends until the end of the 1\,s duration we use in this study. However, their SASI modes have a very low amplitude after $\sim 300$\,ms, which isn't visible to our model at the SNRs considered here. The end time of the SASI mode was measured to be $0.22^{+0.09}_{-0.14}$\,s for model s12, and $0.31^{+0.20}_{-0.08}$\,s for model s14. Although we are not able to capture the entire SASI mode for these models, the reconstructed SASI mode is still longer than for the regular neutrino-driven explosions. 

For model m20, the g/f-mode ends abruptly as the simulation was ended too early to capture the entire gravitational-wave signal. The m20 model is another example of how our CCSN model can reconstruct the g/f-mode produced by a different CCSN simulation code. For the m20 model, we are able to reconstruct a long lower frequency SASI mode. The duration of the SASI mode was measured as $0.34^{+0.04}_{-0.04}$\,s. 

Model s18np is the only model where the SASI mode has a longer duration than the higher frequency g/f-mode. 
Unlike subsequent CoCoNuT models, this model was calculated with a smaller value for the nucleon strangeness, resulting in less optimistic heating conditions more favourable to development of SASI activity.
We measure the duration of the SASI mode to be $0.39^{+0.02}_{-0.45}$\,s. The long SASI mode reconstructions of these non-exploding models are clear indicators that their gravitational-wave progenitor stars likely did not undergo shock revival. 

\subsection{Black Hole Supernovae}

\begin{figure}
\includegraphics[width=0.5\textwidth]{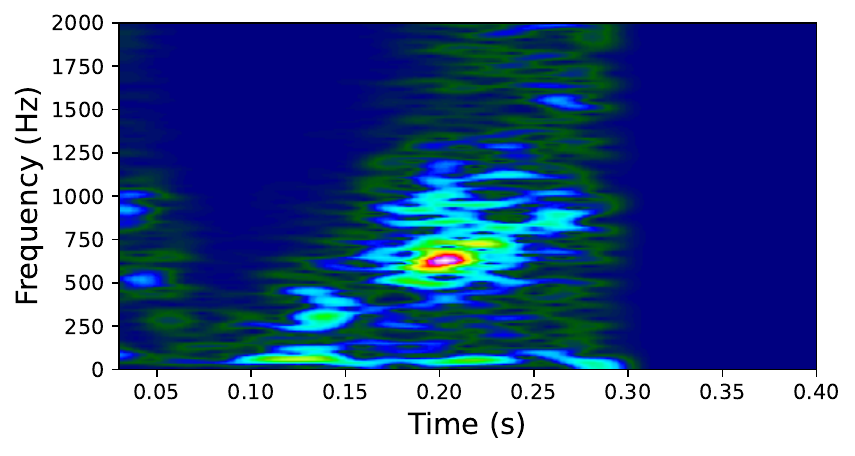}
\includegraphics[width=0.5\textwidth]{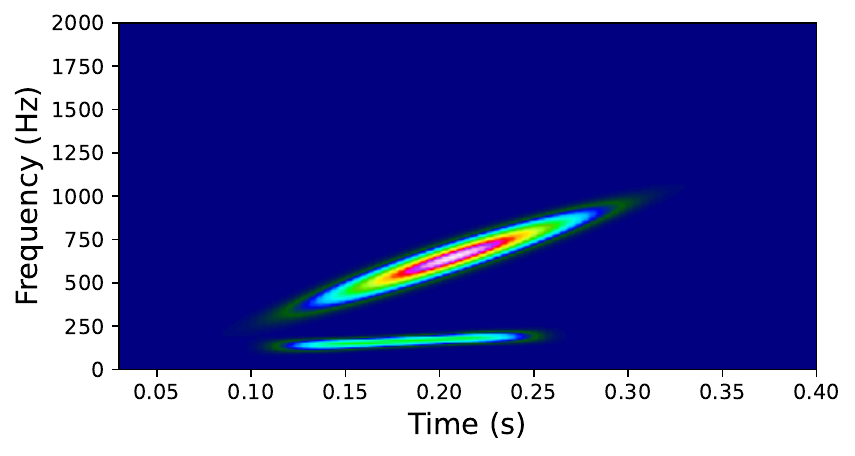}
\includegraphics[width=0.5\textwidth]{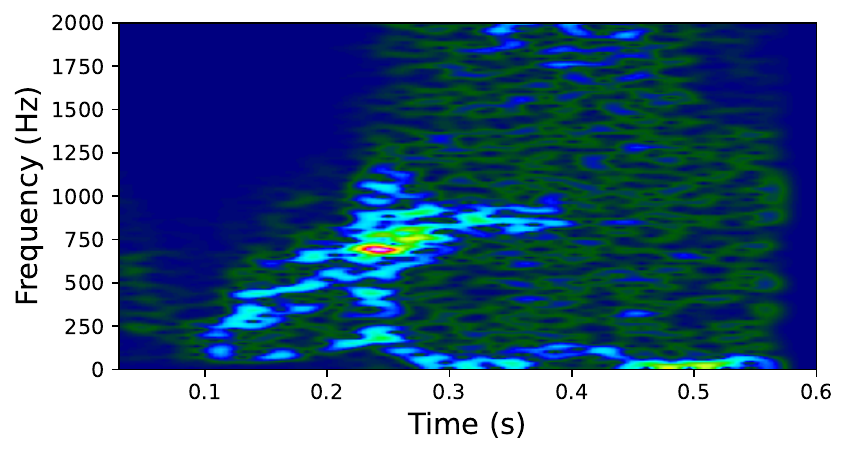}
\includegraphics[width=0.5\textwidth]{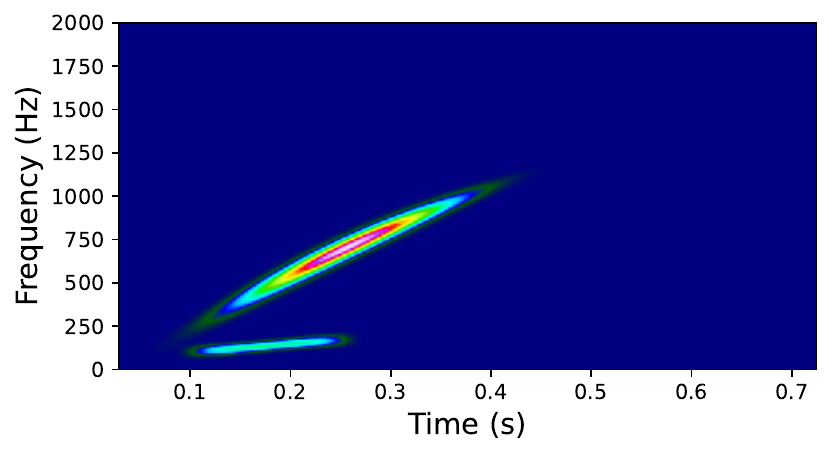}
\includegraphics[width=0.5\textwidth]{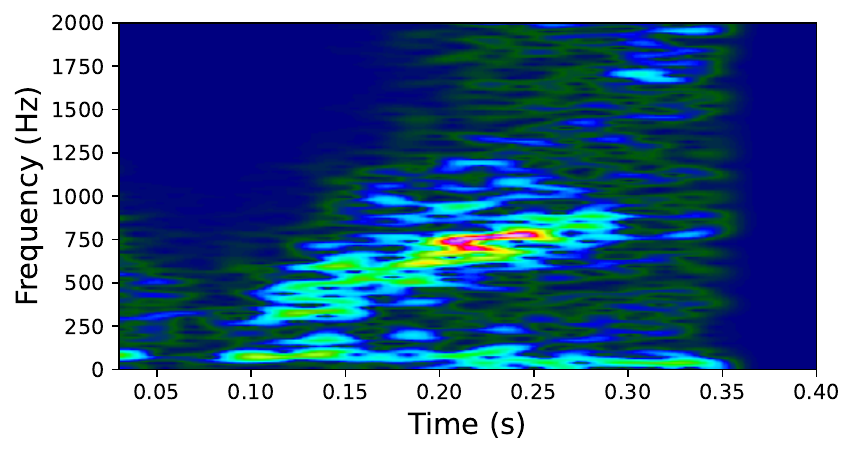}
\includegraphics[width=0.5\textwidth]{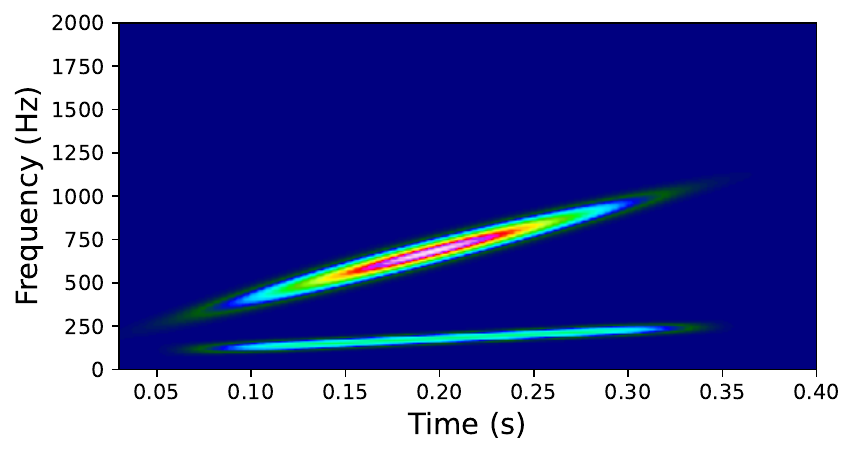}
\caption{Examples of reconstructed gravitational-wave signals for models that undergo shock revival before quickly forming a black hole. We refer to these types of CCSNe as BHSNe. From top to bottom are models z85\_ls220, z85\_SFHx and z85\_SFHo. The left column shows the injected waveforms. The right column shows the gravitational-wave signal as reconstructed by our model at SNR 40. The models have strong SASI modes that increase in frequency with time before the shock is revived.}
\label{fig:bhsn_results}
\end{figure}

Examples of the reconstructed gravitational-wave signals for the BHSNe models are shown in Figure \ref{fig:bhsn_results}. Due to the rapid black hole formation, these models are of much shorter duration than simulations of regular CCSNe and the non-exploding models. The BHSNe models undergo shock revival a few hundred milli-seconds before forming a black hole. The short time between shock revival and black hole formation prevents the shock wave from reaching the star's surface. 

For all three waveform models, we are able to accurately reconstruct the high-frequency g/f-mode at all the SNR values considered. This is the first test of our CCSN signal model for shorter duration CCSN gravitational-wave signals. The g/f-mode increases in frequency with time much more rapidly than for regular CCSNe and the non-exploding models. This is due to a high rate of mass accretion onto the PNS and a rapidly shrinking PNS radius. Therefore, a reconstructed gravitational-wave g/f-mode with a high rate of change of frequency may be another indicator for rapid black hole formation.  

All three injected models contain strong low-frequency gravitational-wave emission before the shock revival. The z85\_ls220 model has very short duration SASI, as the shock revival occurred very early at 0.16\,s post bounce, and the emission is more convection dominated. The z85\_SFHx and z85\_SFHo models have much stronger SASI modes, as they undergo shock revival at 0.3\,s and 0.21\,s respectively. The end times for the SASI modes were measured to be $0.15^{+0..06}_{-0.06}$\,s for z85\_ls220, $0.17^{+0.06}_{-0.04}$\,s for z85\_SFHx and $0.24^{+0.08}_{-0.03}$\,s for z85\_SFHo at SNR 40. The reconstructed SASI modes for z85\_ls220 and z85\_SFHo are longer than expected, because all of these models still have strong low frequency gravitational-wave emission after shock revival. The later shock revival time for model z85\_SFHx results in the SASI mode reaching a higher frequency before shock revival, making it easier for a more accurate reconstruction of the SASI end time. 

The duration of the SASI mode for BHSNe is very similar to the case of regular obscured CCSNe, so the main indicators of black hole formation in these models are the abrupt end of the g/f-mode and the g/f-mode's rapid increase in gravitational-wave frequency. 

\section{CCSN Model Selection}
\label{sec:emdark_model}

\begin{figure}
\centering
\includegraphics[width=0.9\textwidth]{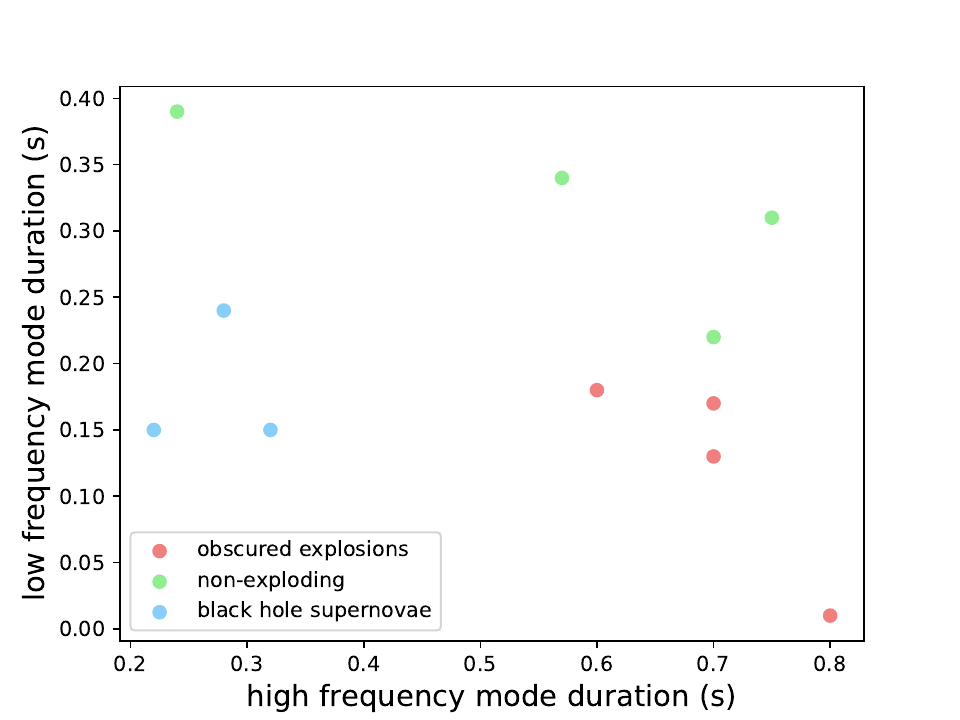}
\caption{The duration of the reconstructed g/f-modes and SASI modes for CCSN models injected at SNR 40. There is a clear distinction between the different types of EM dark CCSNe. The exception is model s18np, which has a very short duration g-mode for a model that does not form a black hole rapidly. }
\label{fig:duration}
\end{figure}

In Figure \ref{fig:duration}, we show the reconstructed duration of the g/f-modes, and the reconstructed duration of the SASI modes for each of the injected CCSN waveform models. Other work to distinguish between different CCSN models has used Bayesian model selection \cite{Powell_17, Powell_23b}, or machine learning techniques \cite{chan_20, Iess_20, saiz_22}. However, the method used here, of simply measuring the mode durations and frequencies, is simpler, more computationally efficient, and similar to work on determining if a black hole forms from the neutrino emission \cite{walk_20}.
The principal component analysis (PCA) models used in \cite{heng_09, logue_12, Powell_17, roma_19, Powell_23b} are also limited by the waveforms used in the PCA model. They would be unable to accurately reconstruct the g/f-mode if the length, or rate of change of frequency, was outside the limits of the values in their training waveforms. 

Figure \ref{fig:duration} shows that the obscured regular CCSN explosions, the non-exploding models, and the BHSNe are clearly separated into different areas of the parameter space. The BHSNe all have g/f-mode durations of below half a second. The obscured regular CCSNe have g/f-mode durations of above 0.5\,s, and SASI mode durations below 0.2\,s. The non-exploding models have both long duration g/f-modes, and long-duration SASI modes. This shows that we can clearly determine the type of EM dark CCSN from a reconstruction of the gravitational-wave signal alone. The exception is model s18np, which could be mistaken for a BHSNe due to its very short g/f-mode. However, the g/f-mode in this model does not rapidly reach high frequencies as we would expect for a BHSNe. Therefore, if we also take into account the rate of change of frequency of the g/f-modes, then we should still be able to accurately determine that s18np is a non-exploding model. It is also expected that the collapse of the PNS to a black hole would create a high amplitude spike of gravitational-wave emission, that may be detectable out to distances of 10\,Mpc \cite{uchida_19}. However, we cannot consider that aspect of the gravitational-wave emission here, as current CCSN simulation codes are not able to calculate it.   

Determining the correct CCSN scenario, for a gravitational-wave signal with no EM counterpart, may also be aided by a coincident detection by neutrino observatories. By the time of O5, currently expected in late 2027, there will be several more sensitive neutrino detectors in operation including Hyper Kamiokande \cite{hyperk_18}, DUNE \cite{dune_20} and JUNO \cite{juno_16}. The SASI also results in a strong imprint on the neutrino signal \cite{tamborra_14,kuroda_17,mueller_19,Zidu_19, Zidu_23, drago_23}. A joint gravitational-wave and neutrino detection would give us increased confidence in the duration of the SASI, and the neutrino detectors would also observe an abrupt end to the neutrino emission in the case of black hole formation. The characteristics of neutrino emission in black hole formation were previously studied by \cite{walk_20} for a $40\,\mathrm{M}_{\odot}$ non-exploding model and a $70\,\mathrm{M}_{\odot}$ BHSNe, but they did not consider the gravitational-wave emission. 

\section{Maximum detectable distances}
\label{sec:dist}

\begin{figure}
\includegraphics[width=\textwidth]{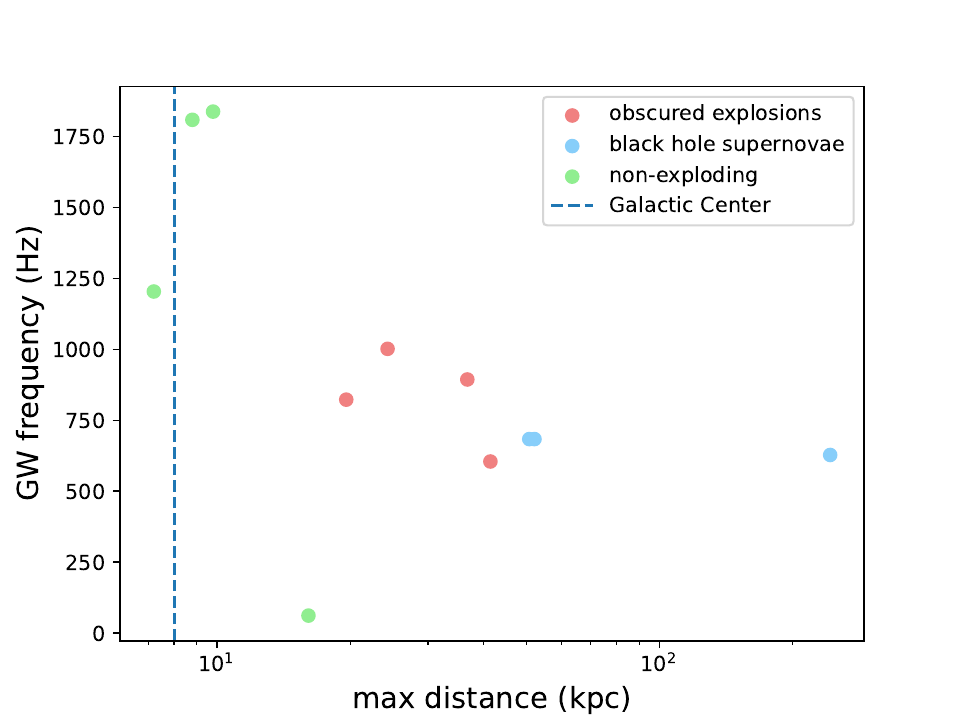}
\caption{The maximum detection distances for CCSNe in a LIGO-Virgo-KAGRA O5 detector network, and the gravitational-wave frequency where the signal amplitude peaks. CCSNe that do not undergo shock revival have the smallest detection distances. Exploding models can be seen out to further distances, and could be obscured by the centre of the Galaxy. BHSNe have larger detection distances due to their higher progenitor masses. CCSNe with rotation and magnetic fields could be seen at much larger distances than the models considered here.
}
\label{fig:distances}
\end{figure}

We consider an estimate of the maximum distance that these different types of CCSN signals may be observable out to in the next LIGO-Virgo-KAGRA observing run O5. The results are shown in Figure \ref{fig:distances}. We consider the maximum detectable distance to be the distance required for an optimal network SNR of 12. This is the maximum possible distance out to which we could see these signals, without accounting for difficulties in developing more sensitive CCSN search algorithms. Although this may be an optimistic SNR for current CCSN gravitational-wave searches, the SNR is still larger than some of the previous gravitational-wave detections \cite{gwtc3}. We mark the Galactic Centre at 8\,kpc. We also include the frequency at which the gravitational-wave amplitude is highest, as the gravitational-wave detectors sensitivity is frequency dependent. 

There is a clear difference between maximum detection distances in O5 between the different CCSN scenarios considered here. For non-exploding CCSNe, they have the lowest gravitational-wave energy and detectable distances. Typical neutrino-driven explosions can be detected throughout the entire Galaxy, and therefore could be obscured behind the Galactic Centre during O5. BHSNe may be detectable beyond the Galaxy as they usually occur in progenitor stars with higher masses. The maximum detection distances would increase if the progenitor stars have rapid rotation or strong magnetic fields, which are both neglected in the models considered in this work. The distance of the CCSN would be unknown if there is no EM counterpart. Measuring the distance would be difficult from the gravitational-wave signal alone.
The gravitational-wave energy is still considered to be quite uncertain, and can vary between different CCSN simulation codes. The distance to a gravitational-wave burst source is typically given by,
\begin{equation}
d^2 = \frac{ E_\mathrm{GW} G }{ h_\mathrm{rss}^2 \pi^2  c^3 f^2 },
\end{equation}
where $E_\mathrm{GW}$ is the gravitational-wave energy, $f$ is the peak frequency, and $h_\mathrm{rss}$ is the root sum squared amplitude of the gravitational-wave signal \cite{sutton_13}. The gravitational-wave amplitude would be measured directly by our signal model, or the CCSN search codes, however some assumption would need to be made about the gravitational-wave energy to determine the distance. Understanding which type of EM dark CCSN occurred may help us to make a reasonable assumption about the gravitational-wave energy.

If neutrinos are observed from the CCSN signal, then it may still be possible to determine the distance to the progenitor star \cite{kachelriess_05, segerlund_21}. In this case, we would be able to convert our measured posterior distributions for the gravitational-wave amplitude into a posterior on $E_\mathrm{GW}$. We have already demonstrated this in our previous work for EM bright CCSNe \cite{Powell_22}. A measurement of $E_\mathrm{GW}$ using the distance from neutrinos would enable us to determine the accuracy of the $E_\mathrm{GW}$ values predicted in CCSN simulations. Knowing the distance and location of the CCSN from the neutrinos, would also aid EM observatories that may want to make follow-up observations. Even when there is no full EM explosion, EM telescopes may still be able to determine if a previously known star has disappeared, or search for the presence of a weaker transient similar to a luminous red nova. 

\section{Discussion and conclusions}
\label{sec:conclusions}

In the 10 years since the first detection of gravitational waves, significant progress has been made in 3D simulations of CCSNe. This has resulted in better gravitational waveforms, with more accurate input astrophysics, and longer waveform durations covering the full explosion phase. Although the gravitational-wave energy of CCSNe is still uncertain, a consensus has been reached in our understanding of the time-frequency morphology of the CCSN gravitational-wave signal. 

The main component of the gravitational-wave signal is now understood to be a g/f-mode that increases in frequency with time as the PNS gains mass and shrinks in radius. In recent years, this understanding has allowed the community to develop universal relations that describe the relationship between the gravitational-wave frequency and the properties of the PNS. In our previous work, we added an asymmetric chirplet model to the Bayesian parameter estimation code Bilby, to enable us to reconstruct the time-frequency morphology of this mode, and convert the results into an estimation of the time evolution of the mass and radius of the PNS.  

A secondary feature expected in the gravitational-wave signal is a lower frequency mode due to the SASI. The SASI mode increases in gravitational-wave frequency with time until the shock is revived, at which point the SASI either disappears, or the gravitational-wave frequency of the SASI mode quickly decreases until it becomes too low for the LIGO-Virgo-KAGRA detectors. 

Searches for gravitational waves from CCSNe have focused on data at times where a CCSN has been observed by EM observatories. However, the first detection of gravitational waves from a CCSN may not have an EM counterpart. This could be due to the explosion being obscured by the Galactic centre, or may occur because the star failed to power a full explosion, or rapidly formed a black hole in a so-called BHSNe. As the SASI mode occurs for a longer duration in failed explosions, previous studies have shown that one can predict if a black hole is formed using the CCSN neutrino signal alone. In this work, we aim to do a similar study with the gravitational-wave emission alone, to determine if the gravitational-wave emission can tell us if the signal was emitted by a CCSN that was obscured, non-exploding or rapidly formed a black hole. 

As a long-duration SASI mode is one of the main indicators of failed explosions, in this work we have adapted a CCSN signal model for g/f-modes to also include a lower frequency SASI mode. 
Previous work has shown that model-agnostic techniques can be used to reconstruct features of the gravitational-wave signal \cite{raza_22, szczepanczyk_21}. However, those techniques are more computationally expensive than what we have shown here, and the extra features they reconstruct may be only related to the stochastic phase of the signal, and may not necessarily  provide you with extra information about the astrophysics of the source over what we can obtain here with our more simplistic reconstruction model. The work of \cite{roma_19} used model selection to determine if the SASI mode is present, but this PCA method cannot determine the length of the SASI mode, which we need to determine if the CCSN undergoes shock revival or forms a black hole. 

We injected 11 different simulated gravitational-wave signals, from the three different CCSN types, into simulated gravitational-wave data at the sensitivity expected during O5. We show how well our model can reconstruct the g/f-mode and SASI mode. We show that obscured CCSNe all have a g/f-mode reconstructed duration $>0.5$\,s, and a SASI mode duration shorter than $0.2$\,s. The non-exploding models also have long duration g/f-modes, but their reconstructed SASI mode is always longer than $0.2$\,s, even if the SNR is too low to capture the entire duration of the SASI mode. For BHSNe, the g/f-mode rapidly reaches high frequencies before ending abruptly. If the gravitational-wave detection was coincident with a neutrino detection, as well as independently confirming the presence of the SASI, the neutrinos may provide a distance estimate, which would allow us to constrain the currently uncertain CCSN gravitational-wave energy.  

In this work, we neglect the impact of magnetic fields and rotation. In 3D CCSN simulations, strong magnetic fields and rapid rotation usually result in rapid explosions with no time to develop a significant SASI component \cite{bugli_20, Powell_23}. The gravitational waves from these explosions would be very similar to the waveforms we consider here for obscured CCSNe, but would be detectable out to larger distances. In some previous work, rotation without strong magnetic fields has resulted in a larger neutrinosphere, less neutrino heating, and therefore later explosion times and longer SASI signatures in the gravitational-wave emission. For CCSNe with a PNS close to the maximum neutron star mass, rotation may also support the PNS from collapsing, leading to a longer timescale to black hole formation \cite{pajkos_25}. However, rapid rotation and strong magnetic fields are thought to occur in only $\sim1\%$ of CCSNe.

\section*{Acknowledgements}
The authors are supported by the Australian Research Council's (ARC) Centre of Excellence for Gravitational Wave Discovery (OzGrav) through project number CE230100016. BM acknowledges support from the ARC through Discovery Project DP240101786. The authors acknowledge computer time allocations from Astronomy Australia Limited's ASTAC scheme, the National Computational Merit Allocation Scheme (NCMAS), and from an Australasian Leadership Computing Grant. Some of this work was performed on the Gadi supercomputer with the assistance of resources and services from the National Computational Infrastructure (NCI), which is supported by the Australian Government, and through support by an Australasian Leadership Computing Grant.  Some of this work was performed on the OzSTAR national facility at Swinburne University of Technology. The OzSTAR program receives funding in part from the Astronomy National Collaborative Research Infrastructure Strategy (NCRIS) allocation provided by the Australian Government, and from the Victorian Higher Education State Investment Fund (VHESIF) provided by the Victorian Government. The authors thank Marco Cavaglia and Simon Stevenson for their helpful comments. 




\bibliography{bibfile}

\end{document}